\documentclass[useAMS,usenatbib]{mn2e}

\usepackage{longtable}
\usepackage{pdflscape}
\usepackage{graphicx}
\usepackage{pdflscape}
\usepackage{lscape,graphicx,pifont}
\usepackage{color}

\newcommand{\Msun}{\mbox{\,M$_\odot$}}
\newcommand{\Lsun}{\mbox{\,L$_\odot$}}

\newcommand{\vunit}{\mbox{\,km\,s$^{-1}$}}
\newcommand{\mic}{\mbox{$\,\mu$m}}
\newcommand{\pion}[2]{{#1}\,{\sc {#2}}}
\newcommand{\fion}[2]{[{#1}\,{\sc {#2}}]}
\newcommand{\nucl}[2]{\mbox{$^{#1}${#2}}}

\newcommand{\ltsimeq}{\raisebox{-0.6ex}{$\,\stackrel
        {\raisebox{-.2ex}{$\textstyle <$}}{\sim}\,$}}
\newcommand{\gtsimeq}{\raisebox{-0.6ex}{$\,\stackrel
        {\raisebox{-.2ex}{$\textstyle >$}}{\sim}\,$}}

\newcommand{\sak}{\mbox{Sakurai's Object}}

\title[Infrared observations of Sakurai's Object]
{V4334~Sgr (Sakurai's Object): still churning out the dust} 

\author[A. Evans et al.]{A. Evans$^{1}$\thanks{E-mail: a.evans@keele.ac.uk},
D. P. K. Banerjee$^{2}$, T. R. Geballe$^3$, 
R. D. Gehrz$^4$, C. E. Woodward$^4$\thanks{Visiting Astronomer at the Infrared
Telescope Facility, which is operated by the University of Hawaii under contract
80HQTR19D0030 with the National Aeronautics and Space Administration.}, \newauthor
K. Hinkle$^5$,
R. R. Joyce$^5$, 
M. Shahbandeh$^6$
\\
$^{1}$Astrophysics Group, Keele University, Keele, Staffordshire, ST5 5BG, UK\\
$^{2}$Physical Research Laboratory, Ahmedabad 380009, India\\ 
$^{3}$Gemini Observatory/NSF's NOIRLab, 670 N. A'ohoku Place, Hilo, Hawai'i, 96720, USA \\
$^{4}$Minnesota Institute for Astrophysics, School of Physics \& Astronomy,
116 Church Street SE, University of Minnesota, \\ Minneapolis, MN 55455, USA\\ 
$^{5}$National Optical-Infrared Astronomy Research Laboratory, 
950 N. Cherry Avenue, Tucson, AZ 85719, USA \\ 
$^{6}$Department of Physics, Florida State University, 77 Chieftain Way, Tallahassee, FL 32306-4350, USA
}

\begin{document}

\date{Version of \today}

\pagerange{\pageref{firstpage}--\pageref{lastpage}} \pubyear{2021}

\maketitle

\label{firstpage}

\begin{abstract}
We present a 0.8--2.5\mic\ spectrum of the Very Late Thermal Pulse 
object V4334~Sgr (Sakurai's Object), obtained in 2020 September. 
The spectrum displays a continuum that rises strongly to longer
wavelengths, and is considerably brighter than the most recent 
published spectrum obtained seven years earlier.
At the longer wavelengths the continuum is well fitted 
by a blackbody with a temperature of $624\pm8$~K. 
However, there is excess continuum at the
shortest wavelengths that we interpret as being due to hot 
dust that has very recently formed in an environment
with C/O $\simeq2.5$. Other possible sources for this excess 
continuum are discussed -- such as the stellar photosphere dimly
seen through the dust shell, and light scattered off the inner wall
of the dust torus -- but these interpretations seem unlikely.
Numerous emission lines are present, 
including those of \pion{He}{i}, \pion{C}{i}, \fion{C}{i}, and 
\pion{O}{i}. Our observations confirm that emission in the \pion{He}{i} 
1.083\mic\ and \fion{C}{i} 0.9827/0.9852\mic\ lines is spatially
extended. The \fion{C}{i} line fluxes suggest that the electron 
density increased by an order of magnitude between 2013 and 2020, 
and that these two lines may soon disappear from the spectrum. 
The flux ratio of the 1.083\mic\ and 2.058\mic\
\pion{He}{i} lines is consistent with the 
previously-assumed interstellar extinction.
The stellar photosphere remains elusive, and 
the central star may not be as hot
as suggested by current evolutionary models.
\end{abstract}


\section{Introduction}
V4334~Sgr (\sak; hereafter SO) has been widely 
considered to be the product of a Very Late Thermal 
Pulse (VLTP) in a low ($\sim$~solar) mass star. The star became 
carbon-rich in mid-1996 \citep{eyres98}, and in late 1997 it ejected a 
carbon-rich dust shell that became optically thick in mid-1998; 
by the end of that year the dust had completely 
obscured the central star \citep[see Figure~2 of][]{durbeck00}. 

As of late 2021, the star has not reappeared. However, observations
during the last $1\frac{1}{2}$ decades \citep{vanhoof07,vanhoof08,hinkle14,vanhoof15a,
vanhoof15b,vanhoof18,hinkle20} have shown that the optical
and near-infrared (NIR) spectra originating inside the dust shell are 
starting to become detectable.

A comprehensive account of the infrared 
(IR) emission from SO was given by \cite{evans20}, who also 
reviewed determinations of the distance and interstellar reddening;
we assume their values here, namely $D=3.8$~kpc and $E(B-V)=0.62$.
\cite{chesneau09} showed that the dust around SO takes 
the form of a disc/torus (hereafter ``disc''). The presence of 
an optically thick disc, the opacity of which is unknown,
clearly renders the determination of the {\em total} 
(interstellar plus circumstellar) reddening somewhat uncertain.
This is discussed below.

Here we present a 0.8--2.5\mic\ spectrum of SO, obtained at the 
Frederick C. Gillett Gemini North Telescope. The new spectrum indicates 
that the trends reported since 2007 are continuing.  

\section{Observations and Data Reduction}

\begin{figure}
 \centering
 \includegraphics[width=8.5cm]{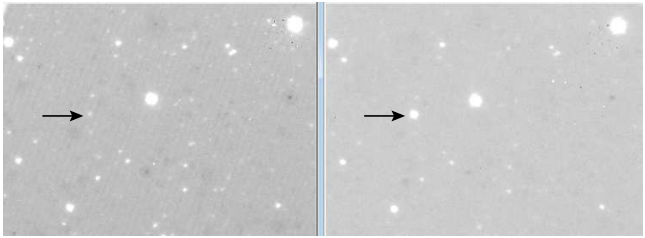}
 \caption{$J\!K$ images of SO obtained on the 3.2~m NASA
 Infrared Telescope Facility on 2020 June 07.46 UT.
 Left: $J$-band image, right: $K$-band image. 
Scale is $\sim45''\times30''$. North is up, east is left.  
SO is arrowed. \label{Sak_JK}}
\end{figure}

\begin{figure}
\setlength{\unitlength}{1mm}
\begin{center}
  \leavevmode
  \includegraphics[width=8cm]{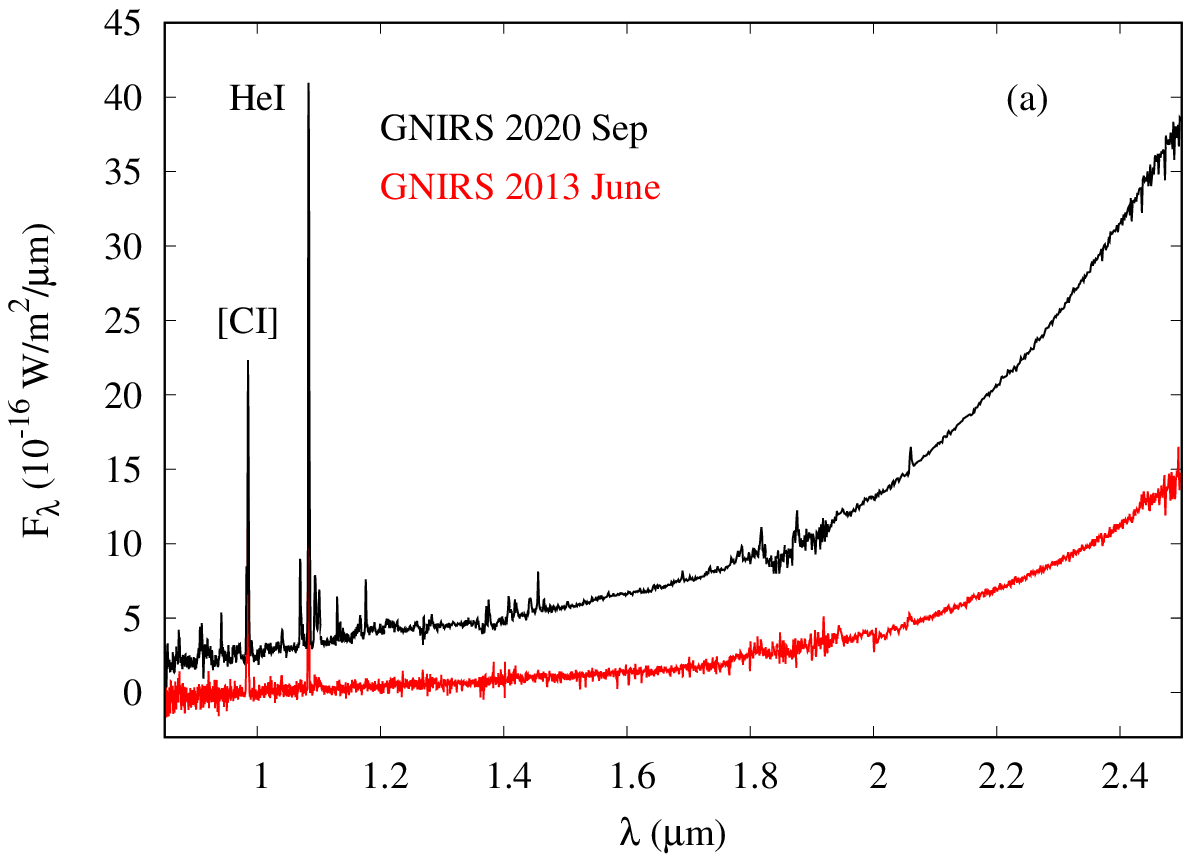}
   \includegraphics[width=8cm]{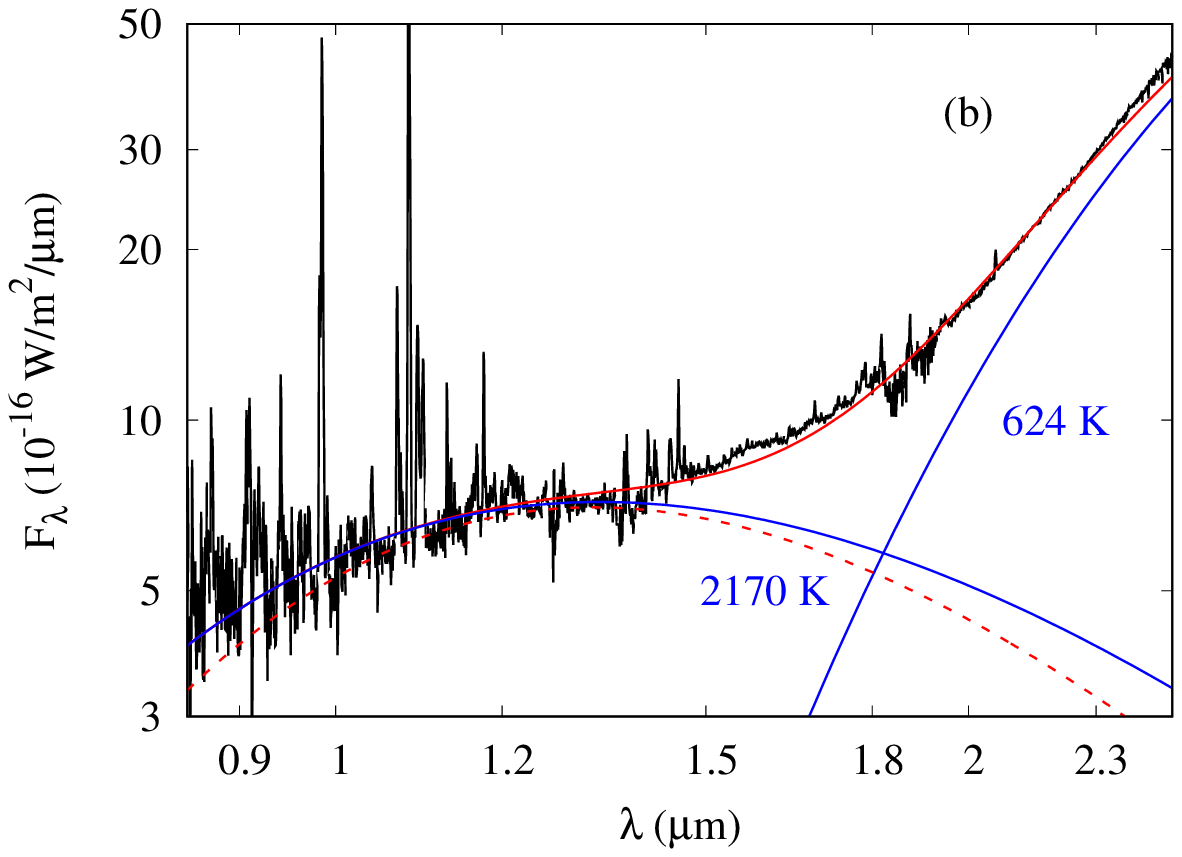}
   \caption[]{(a)~Comparison  of the observed 2013 June spectrum 
   \citep[red;][]{hinkle14} 
   and the 2020 September spectrum (black). (b)~The 2020 spectrum (black), 
   dereddened by $E(B-V)=0.62$; wavelength scale is logarithmic to stretch 
   the scale at the shortest wavelengths.
   The full red curve is a two-blackbody fit to the continuum; the blue 
   curves are the individual blackbodies at the temperatures indicated.
   The dotted red curve is a $10^4$\Lsun, $8\times10^4$~K blackbody,
   reddened by $A_V=9.3$~mag, as described in the text.
   \label{spec}}
\end{center}
\end{figure}

\begin{figure*}
\setlength{\unitlength}{1mm}
\begin{center}
  \leavevmode
   \includegraphics[width=0.8\textwidth]{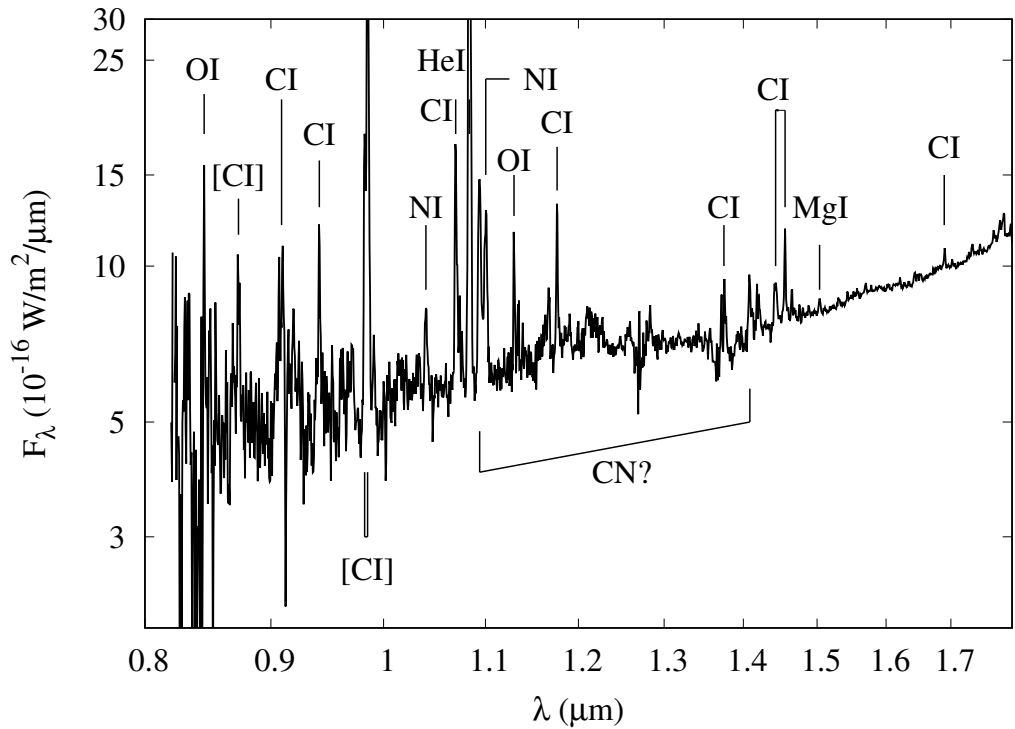}
   \caption[]{Identification of emission lines in the 2020 September
   spectrum; see also Table~\ref{fluxes} and Fig.~\ref{CN_1000}. 
   Wavelength scale is logarithmic to stretch the scale at the 
   shortest wavelengths. The small dips in the apparent continuum 
   near 1.27\mic\ and in the 1.35--1.40\mic\ interval are due to 
   incomplete cancellation of strong telluric O and H$_2$O absorption 
   bands, respectively. The emission features in that interval are 
   real and intrinsic to SO.   
   \label{lines}}
\end{center}
\end{figure*}

Spectra of SO and a telluric standard, HIP~90404 (A0V), were  
obtained on UT 2020 September 29  for 
programme GN-2020B-FT-206. The facility NIR spectrograph GNIRS 
\citep{elias06} was used in its cross-dispersed mode 
with a $0\farcs45\times7\farcs0$ slit, yielding a spectral resolving power, 
$R$,  of $\sim1200$ across the 1.0--2.5\mic\ band, and slightly lower $R$ 
at shorter wavelengths. The standard nod-along slit (ABBA) mode was used 
with a nod of $\pm1\farcs5$ from slit centre. The slit was oriented at a 
position angle of $35^\circ$ east of north. The total exposure 
time on SO was 2400~seconds. SO and the telluric standard
were observed at airmasses 
within several hundredths of 1.4. The spectra were obtained in clear 
skies with good and stable seeing. Measurements of the spectral images 
yield full widths at half maxima of 0\farcs57 and 0\farcs51, at 1.0\mic\
and 2.2\mic, respectively.

Data reduction with the Gemini {\sc iraf} \citep{iraf} 
and Starlink {\sc figaro} \citep{figaro} packages involved the 
standard steps of flat-fielding, spatial and spectral rectification 
of the images, removal of the effects of cosmic ray hits, order by 
order extraction of the spectra, spectral cross correlation and 
combining of the negative and positive spectra, and wavelength 
calibration using an argon lamp. Following the removal of most of the 
photospheric hydrogen recombination lines in the spectrum of 
the telluric standard and wavelength alignment of the SO
and telluric standard, flux calibration was achieved by ratioing 
the two spectra under the assumption that HIP~90404 has the 
continuum of a 9480~K blackbody and the $J\!H\!K$ magnitudes 
listed in SIMBAD\footnote{http://simbad.u-strasbg.fr/simbad/}. 
We estimate that the flux calibration is accurate to $\pm20$\%. 

The $J$ and $K$ band images shown in Fig.~\ref{Sak_JK} were 
obtained with the SpeX \citep{rayner03} guider imager on the 
NASA Infrared Telescope Facility on 2020 June 07.46 UT, 
using a five point dither with 40~s exposure time at each position, 
and at an average airmass of 1.26. Data reduction was done in a 
standard manner, involving sky and  dark subtraction, and flatfielding. 
Instrumental magnitudes obtained through aperture photometry were 
calibrated against several 2MASS field stars; we find 
$J = 16.89\pm0.09$~mag, $K_s = 12.66\pm0.22$~mag for SO at this epoch.

\section{Results and discussion}

The observed spectrum, extracted over a $1''$ region of the slit
centred on the continuum peak, is shown in  Fig.~\ref{spec}a, along with 
the spectrum obtained by \cite{hinkle14} with GNIRS on 2013 June 14. 
The comparison indicates that significant changes occurred in the 
intervening $\sim7$~years. In our spectrum, the apparent emission 
features around 1.8--1.9\mic\ are due to incomplete removal of the 
hydrogen absorption features in the spectrum of the telluric standard. 

Fig.~\ref{spec}b shows the new spectrum, along with blackbody fits 
to the continuum, which are discussed in Section~\ref{BB}.
Numerous emission lines are present, especially in the 
0.8--1.7\mic\ region. This spectral interval is shown in 
more detail in Fig.~\ref{lines}, where most of the lines are identified. 

\subsection{The continuum\label{BB}}

As is evident from Fig.~\ref{spec}a,b, the flux density was rising 
from $\sim1.4$\mic\ to longer wavelengths in 2013 and 2020, indicating 
emission by dust. In this respect the 2013 and 2020 spectra are 
broadly similar. However there are two key differences:
First, in 2020 there is an additional excess above the 
dust continuum at the shortest wavelengths: no 
continuum around 1\mic\ was detected by \cite{hinkle14}; its 
contribution was $\ltsimeq10^{-16}$~W~m$^{-2}$\mic$^{-1}$ at 1.4\mic.
Second, the continuum flux density was considerably greater at all 
wavelengths in 2020; for example it was three times higher at 2.4\mic.

The spectra and photometry indicate a  
steady rise in the flux in the $K_s$ 
band, from $K_s>18.4$ in 2009 March, to $K_s\simeq14$ around the 
time of the \citeauthor{hinkle14} observation, to $13.20\pm0.02$ 
in 2017 October \citep[see Table~2 and Figure~3 in][]{evans20}, 
to $K_s=12.66\pm0.22$, close to the time of the 2020 September
GNIRS observation. 

We assume that the NIR
continuum is subject to interstellar extinction only;
we briefly explore the effects of internal extinction below.
A fit of two blackbodies to the continuum from the 2020 spectrum,
dereddened by $E(B-V)=0.62$ \citep[][see also below]{evans20},
gives temperatures of $624\pm8$~K (``Component~1'') and 
$2170\pm25$~K (``Component~2'') for the two components
(see Fig.~\ref{spec}b). 
Component~1 is clearly due to emission 
by dust. Assuming that the dominant dust component is 
amorphous carbon \citep[AC; see][and references therein for a 
justification of this]{evans20}, we deduce a dust mass of 
$5.7[\pm1.6]\times10^{-10}$\Msun\ and a bolometric dust 
luminosty of $38[\pm10]$\Lsun\ for a distance $D=3.8$~kpc.

\cite{hinkle14} determined that, in 2013, the dust continuum 
(corresponding to our Component~1) had a temperature of 590~K;
a recalculation using the same procedure as that used here
(i.e. same dereddening, same blackbody fit), for consistency, 
yields $550\pm40$~K; the corresponding dust mass and luminosity 
are $8.5[\pm3.8]\times10^{-10}$\Msun\ and 31\Lsun\ respectively. 
There is no significant difference between
the dust masses in 2013 and 2020.

The interpretation of Component~2 is not as straight-forward.
The value of 
$[\lambda{f}_\lambda]_{\rm max}$ for this component is 
$1.07\times10^{-15}$~W~m$^{-2}$, or $\sim0.5$\Lsun\ at 3.8~kpc. 
There appear to be three possibilities, which we discuss below: 
(a)~the stellar photosphere is being viewed, through several 
magnitudes of visual circumstellar extinction, 
(b)~light from an embedded stellar remnant
is scattered off the inner wall of the dust disc,
(c)~the emission is from extremely 
hot, very recently-formed dust.

\paragraph*{(a) The stellar photosphere.}

Evolutionary models \citep[e.g.,][]{hajduk05} suggest that, in its 
current ($\sim2021.0$) state, SO should have had an effective 
temperature $T_*\sim10^5$~K and bolometric luminosity 
$L_*\sim10^{3.8}$\Lsun. Such a source, at a distance of 3.8~kpc,
would have 
$[\lambda\,f_\lambda]_{\rm max}\simeq2.2\times10^{-11}$~W~m$^{-2}$,
but would be seen through a large (but unknown) amount of circumstellar
extinction and reddening (parametrised by the visual extinction, 
$A_V$, in magnitudes). A reddening law in which
$A_\lambda$ is approximately proportional to $\lambda^{-1}$ has 
the effect of making a blackbody appear cooler than it actually is, 
and for large reddening a $10^5$~K blackbody can appear as cool 
as $\sim3000$~K. 

\begin{landscape}
 
\begin{table*}
\caption{Emission lines in the spectrum of SO; the
spectrum was extracted over a $1''$ region of the slit, 
centered on the continuum peak.
$\lambda_{\rm obs}$ and $\lambda_{\rm id}$ are the measured and 
listed vacuum swavelengths respectively, the latter from van Hoof (2018) unless 
specified otherwise. Dereddening in the penultimate column is by $E(B-V)=0.62$. 
\label{fluxes}}
 \begin{tabular}{lcccrrl} 
 &  &  & Transition &  \multicolumn{2}{c}{Line flux ($10^{-18}$~W~m$^{-2}$)} &\\ \cline{5-6} 
  \multicolumn{1}{c}{$\lambda_{\rm obs}$ ($\mu$m)}  & Identification  & $\lambda_{\rm id}$ ($\mu$m)$^*$ & $u-{\ell}$ & \multicolumn{1}{c}{Observed} & \multicolumn{1}{c}{Dereddened} & \multicolumn{1}{c}{Comment}\\ \hline
0.8455 & \pion{O}{i} &  0.8448 & $^3$P$-^3$S$^o$ & $0.45\pm0.04$ & $1.11\pm0.10$ & \\
0.8732 & \fion{C}{i} &  0.8729 & $^1$S$_0-^1$D$_2$ &$0.44\pm0.05$ & $1.09\pm0.14$ & Wavelength from \cite{haris17}.\\
0.9090 & \pion{C}{i} &  0.9094--0.9097 & $^3$P$-^3$P$^o$ & $1.13\pm0.13$ & $2.32\pm0.25$& Blend of 5 lines. \\ 
0.9417 & \pion{C}{i} &  0.9408  & $^1$D$_2-^1$P$^o_2$  &    $0.50\pm0.05$ & $1.03\pm0.12$  & \\
0.9825 & \fion{C}{i} & 0.9824 &  $^1$D$_2-^3$P$_1$ & $0.90\pm0.05$ & $1.92\pm0.12$ & Wavelength from \cite{haris17}.\\
       &&&&&& Detected by \cite{hinkle14}. \\
0.9852 & \fion{C}{i} & 0.9850 &  $^1$D$_2-^3$P$_2$ & $4.00\pm0.07$ & $8.45\pm0.16$ & Wavelength from \cite{haris17}\\
&&&&&& Detected by \cite{hinkle14} \\
1.0405 & \pion{N}{i} & 1.0405 & $^2$D$_{3/2}-^2$D$^o_{5/2}$ &  $0.30\pm0.03$   & $0.63\pm0.05$ & Detected by \cite{hinkle14}.\\
       &              & 1.0411 & $^2$D$_{5/2}-^2$D$^o_{5/2}$ &    &  & \\

1.0698 & \pion{C}{i} & 1.0688 & $^3$D$-^3$P$^o$ & $1.29\pm0.13$ & $2.47\pm0.09$ & \\ 
1.0835 & \pion{He}{i}& 1.0833 & $^3$P$^o-^3$S & $8.94\pm0.12$  & $16.88\pm0.24$ & Detected by \cite{hinkle14}.\\
1.0940 &  CN?       & & & $1.53\pm0.14$ & $2.86\pm0.07$ & Detected by \cite{hinkle14} but unidentified. \\
       &            & & &               &              & See Section~\ref{mol}\\
1.1003 &  \pion{N}{i} & 1.0992  &  $^2$F$_{7/2}-^2$F$^o_{7/2}$  & $1.23\pm0.15$ & $2.31\pm0.07$ & Detected by \cite{hinkle14} but unidentified. \\
1.1298 & \pion{O}{i} & 1.1290 & $^3$D$^o_3-^3$P$_2$ & $0.44\pm0.03$ & $0.80\pm0.05$ & \\
1.1671 & \pion{C}{i} & 1.1656--1.1677 & & $0.34\pm0.05$ & $0.42\pm0.06$ & Blend of 6 lines.\\
1.1762 & \pion{C}{i} & 1.1751--1.1781  & & $0.65\pm0.03$ & $1.11\pm0.06$ & Blend of 4 lines.\\
1.3756 &  \pion{C}{i}  & 1.3727  & $^1$S$_0-^1$P$^o_1$   & $0.22\pm0.02$ & $0.34\pm0.04$ & \\
1.4085 &  CN?             & & & $0.36\pm0.04$ & $0.88\pm0.09$ & See Section~\ref{mol}\\
1.4429 & \pion{C}{i} &  1.4424   & $^3$D$^o_3-^3$P$_2$  & $0.43\pm0.04$  & $0.68\pm0.06$ & \\
1.4557 & \pion{C}{i} & 1.4547   & $^1$P$_1-^1$P$^o_1$  & $0.54\pm0.03$ & $0.81\pm0.05$ & \\
1.4650 & \pion{C}{i}]& 1.4641 &  $^3$F$_{3}-^3$P$^o_{2}$  & $0.15\pm0.03$ & $0.22\pm0.03$ & \\
1.5036 & \pion{Mg}{i}&  1.5029--1.5052& $^3$S$^o-^3$P & $0.10\pm0.01$  & $0.17\pm0.02$ & Blend of 3 lines.\\
1.6902 & \pion{C}{i}  & 1.6895 & $^1$F$^o_3-^1$D$_2$ &   $0.19\pm0.02$  & $0.24\pm0.03$          &  \\
2.0602 & \pion{He}{i}& 2.0589 & $^3$P$^o-^3$S  &$0.45\pm0.03$ & $0.53\pm0.04$ & 
Detected by \cite{hinkle14}.\\ 
&&&&&& {Weak P-Cygni profile?} \\\hline
\multicolumn{7}{l}{$^*$Where a line is identified as a ``Blend'' the range of wavelengths is given.}\\
  \end{tabular}
\end{table*}

\end{landscape}

We have explored the [$L_*, T_*, A_V$] parameter space, and guided by 
the values in \citeauthor{hajduk05} for $L_*$ and $T_*$ and 
reasonable values of $A_V$; for example, \cite{tyne02} determined that
$A_V$ was in the range 8--12 in the period 1999--2001. 
The reddening law for the circumstellar dust is of 
course unknown; we assume that it is
interstellar-like in the 0.8--2.6\mic\ wavelength range.
We find that a $L_*=10^4$\Lsun, $T_*=8\times10^4$~K blackbody, 
reddened by an amount corresponding to $A_V=9.3$, is satisfactorily
similar to Component~2 (see Fig.~\ref{spec}b). However such a
source would easily have been detected by \cite{hinkle14} in 
2013 June, even allowing for 7-year
variations in the stellar and dust parameters. This interpretation
seems therefore unlikely.

\paragraph*{(b) Scattered light.}
Might Component~2 be due to light from the
obscured star, scattered off the inner wall of the disc? 
\cite{chesneau09} show that the inclination of the 
disc is $\simeq75^\circ$, and its large scale height 
($\sim50$~{au}, scaled to $D=3.8$~kpc)
is such that it limits the opening angle at the ``poles''; they
also determine the dimensions of the disc which, scaled to 3.8~kpc, 
are $\simeq115\times150$~{au}.

The scattering of light of a star
embedded in an IR reflection nebula has been considered by 
\cite{holbrook98} and \cite*{pendleton90}, and we primarily
use the latter approach here. The flux from Component~2 in 2020 was 
$\sim10^{-15}$~W~m$^{-2}\mic^{-1}$. With a $0\farcs45$ slit,
and possible extended emission over $\sim1''$ (see 
Section~\ref{ext0} below), this corresponds to an intensity
$I_\lambda\gtsimeq11.4\times10^{-5}$~W~m$^{-2}\mic^{-1}$~sr$^{-1}$, 
the lower limit arising because the extent of the scattered light
may be less than that given in Section~\ref{ext0}.
Using the formulae in \citeauthor{holbrook98} and
\citeauthor{pendleton90}, we get
\begin{eqnarray}
I_\lambda & = & \left ( \frac{L_*}{4\pi\sigma{T}_*^4}\right )
         \frac{B_\lambda(T_*)}{r^2} \:\: \frac{\omega}{4} \:\: \Delta_{\rm thin} 
               \nonumber \\
  & = & \left ( \frac{L_*}{4\pi\sigma{T}_*^4}\right )
  \frac{B_\lambda(T_*)}{r^2} \:\: \frac{\omega}{4} \:\: \Delta_{\rm thick}  
        \label{scat}
\end{eqnarray}
for the optically thin and optically thick cases respectively, where
``optically thin'' and ``optically thick'' refer to the 
optical depth $\tau$ of the scattering layer. $L_*$ and $T_*$ are the 
luminosity and temperature of the illuminating star respectively,
and $r\simeq120$~{au} is the distance 
of the star from the scattering surface.
$\Delta_{\rm thin} = \tau/\cos\theta$ and 
$\Delta_{\rm thick}=\cos\theta_0/(\cos\theta+\cos\theta_0)$,
where $\theta$ ($\theta_0$) is the angle of scattering (incidence).
We take optical constants for AC grains from \cite{hanner88};
the albedo $\omega\sim0.019$ for 0.1\mic\ AC grains at 1.3\mic\
(the wavelength of maximum intensity).

The luminosity of the cooler dust seen in 
SOFIA data in mid-2016 was $\sim3000$\Lsun\ \citep{evans20}. 
There are no long wavelength data later than 2016 but the 
time-dependence of the dust luminosity \citep{evans20} suggests 
that it would have been of this order around the time of the 2013 
and 2020 NIR spectra. The 3000\Lsun\ component must be radiation 
from the central star, reprocessed by the circumstellar dust; 
this value therefore serves as an estimate of the luminosity 
of the embedded star.

We have no value for $T_*$, but the function 
$B_\lambda(T_*)/T_*^4$ has a maximum value of 
$9.98\times10^{-9}$~W~m$^{-2}~\mic^{-1}$~K$^{-4}$ at 1.3\mic, irrespective
of the value of $T_*$. Combining this with the lower limit on
$I_\lambda$, we obtain a lower limit $\Delta\gtsimeq5.2\times10^{-4}$.
We have no information about the
scattering geometry, or of the optical depth $\tau$ to the scattered 
light through the dust shell. But surely this lower limit on 
$\Delta$ can be satisfied by a wide range of scattering 
geometries and optical depths. If Component~2 is scattered
light then its intensity seems consistent with the presence
of an embedded  source, of unspecified temperature.

However Component~2 was not present in 2013.
From Equation~(\ref{scat}), the scattered intensity is 
$\propto{L_*}$. The dimensions of the disc are such that any 
scattered light would, by virtue of the finite speed of light, 
lag behind variations in the star by $\sim1.5$~days, so the 
stellar and scattered light variations would essentially be in phase.
The absence of Component~2 in 2013 would therefore have to
be because either
(i)~the central star was much fainter 
in 2013 than it was in 2020, by a factor $\gtsimeq10$ 
(the flux ratio 2020/2013), (ii)~the light reaching
the scattering surface from the (unchanged) star was 
reduced by internal extinction, or (iii)~the scattered light 
is itself obscured by internal extinction {\em en route} to the
observer.

Case~(i) would require that the 
luminosity of the star in 2013 would 
need to have been $\ltsimeq300$\Lsun, but the dust luminosity 
in 2013 is known to be $\gtsimeq3500$\Lsun\ \citep{evans20}, 
which sets a lower limit on the luminosity of the embedded star.
This inconsistency 
would lead us to conclude 
that this interpetation is unlikely.

However, given SO's proclivity for
continuous dust ejection \citep{evans20}, it it not inconceivable
that a discrete dust cloud \citep[such as those described by][]{hinkle20}
happened to have been ejected around 
2013 --- away from our line-of-sight --- so that light from 
the central star was prevented from reaching the scattering 
surface. Therefore, the observed scattered light would
correspondingly be reduced. This behaviour is similar to
that of the R~Coronae Borealis stars during their well-known
``dust dips'' \citep{clayton12}. Such a cloud might also obscure
the {\em scattered} light even if the star itself is visible 
from the scattering surface. These two scenarios refer respectively 
to cases~(ii) and (iii) above.

In case~(ii), the cloud must 
completely hide the star from the scattering surface, and 
its dimensions perpendicular to the light path must 
therefore be comparable with the disc 
scale height.

We assume a spherical 
cloud of radius $\ell$, where $\ell$ is a few {au},
and estimate the likely properties of the 
putative dust cloud as follows. Comparing the 2013 and 2020
fluxes at 1.25\mic\ (where there are no prominent spectral 
features) we estimate the optical depth in the cloud to be 
$\tau_{1.25}\simeq2.1$, assuming that the underlying
continuum source has not varied between 2013 and 2020.
Taking optical constants for 
AC grains from \cite{hanner88}, we determine the extinction 
efficiency $Q_{\rm ext}$ at 1.25\mic\ to be 0.253. The 
optical depth, together with the extinction efficiency, gives 
the number of grains per unit volume, $n$, in the cloud as
\[ n = \tau_{1.25}/[\pi{a}^2Q_{\rm ext}\:2\ell]] \:\:,\]
where $a$ is the grain radius and $2\ell$ is the path length
through the cloud.

The mass of dust in the obscuring cloud  is 
\begin{eqnarray*}
 M_{\rm dust} & \sim & n \: \frac{4\pi{\ell^3}}{3} \: \frac{4\pi{a}^3\rho}{3} 
 = \frac{8\pi{a}\rho}{9} \:\: \ell^2 \:\: 
  \frac{\tau_{1.25}}{Q_{\rm ext}} \\
 & \simeq & 4.0\times10^{-11} \:\: \left ( \frac{\ell}{\mbox{au}}
 \right)^2  \:\: \left ( \frac{a}{0.1\mic} \right ) \:\: \Msun\:\:,
\end{eqnarray*}
where $\rho$ is the density of the grain material 
\citep[taken here to be 1.5~gm~cm$^{-3}$, 
the value assumed in][although it transpires that the crucial
result is independent of $\rho$]{evans20}. The temperature of the
dust in the cloud, assuming AC and the data 
in Appendix~B1 of \cite{evans17}, is
\[ T_{\rm dust} \mbox{(K)}\simeq 540 \:\: \left \{
\left ( \frac{L_*}{3350\Lsun} \right )\: 
\left ( \frac{a}{0.1\mic} \right )\:
\right \}^{1/(\beta+4)}  \]
for a grain that is half-way between 
the star and the scattering surface. The parameter
$\beta$ is defined such that the dust emissivity is
$Q_{\rm abs}\propto\lambda^{-\beta}$; for AC,
$\beta=0.754$ \citep[see][]{evans17}. The reference value 
of $L_*=3350$\Lsun\ is from \cite{evans20} for 2014, 
the nearest datum to 2013. Radiation from such a cloud 
would have $[\lambda{f}_\lambda]_{\rm max}$
at $\sim6.8$\mic.

The dust mass can be converted to 
$[\lambda{f}_\lambda]_{\rm max}$ for the obscuring dust cloud 
using the formulae in \cite{evans17}. The quantity
$[\lambda{f}_\lambda]_{\rm max}$ is independent of grain size, 
temperature and density:
\[  [\lambda{f}_\lambda]_{\rm max}  =  \frac{L_*}{32.62\pi^2D^2} \:\:
  \frac{\tau_{1.25}}{Q_{\rm ext}} \left ( \frac{\ell}{r} \right )^2 \:\:.\]
Thus
\[ [\lambda{f}_\lambda]_{\rm max} \simeq 2.45\times10^{-12} \:
\left ( \frac{L_*}{3350\Lsun} \right )\: 
\left ( \frac{\ell}{1~\mbox{au}} \right )^2 \:
\left ( \frac{r}{\mbox{{au}}} \right )^{-2} \:,\]
where $[\lambda{f}_\lambda]_{\rm max}$ is in W~m$^{-2}$.

Even a modest $\ell=5$~{au} 
($\ll$ the disc scale height) and $r\simeq50$~{au} 
(about half-way from the star to the scattering surface) gives 
$[\lambda{f}_\lambda]_{\rm max}\simeq2.44\times10^{-14}$~W~m$^{-2}$,
significantly larger than the observed $1.07\times10^{-15}$~W~m$^{-2}$
for Component~2. Hence, emission from the dust cloud would
overwhelm the fluxes in Fig.~\ref{spec}. The same conclusion 
applies to a dust cloud that obscures the {\em scattered}
(as opposed to the direct) light, although in this case the 
dust temperature would be somewhat cooler.

We argue that Component~2 in 2020, and its 
absence in 2013, was the result of neither an intrinsic 
fading of the star (case~(i)),the extinguished starlight 
scattered off the inner wall of the disc (case~(ii)), nor 
the effect of a dust cloud preventing scattered 
starlight from reaching the observer (case~(iii)).

\paragraph*{(c) Hot dust.} 
We consider whether the component with blackbody 
temperature 2170~K is {\em extremely hot}, freshly-formed, 
dust. While this temperature 
may appear high for dust, it is not unusual for carbon dust.  
For example \cite{gall14} found $T=2300$~K 
on day~26 in supernova SN 2010jl, 
while hot dust has been found in novae 
(e.g. $\gtsimeq1400$~K in V2362~Cyg \citep{lynch08} 
263 days after outburst, and 2000~K in V838~Her
\citep{harrison94}. Also, it is well-known that 
very small ($\sim10$\,\AA) carbon grains (such as those that are
newly formed) can attain very high temperatures by stochastic
heating by single ultra-violet photons \citep{sellgren84}. 
There was no silicon carbide feature at 11.5\mic\ in the mid-IR 
spectrum of SO \citep{evans20}, so this hot component is 
likely to be carbonaceous rather than SiC. 

If, as seems probable, the dust is AC, then 
its actual temperature is somewhat less than 2170~K. For a grain 
with $\beta=0.754$, the wavelength of maximum $f_\lambda$ 
is given by $\lambda_{\rm max}T=2890\times5/(\beta+5)$~\mic-K
rather than $\lambda_{\rm max}T=2890$~\mic-K. The flux from
Component~2 peaks at $\sim1.3$\mic, so that the dust temperature
is 1880~K. This is clearly in line with hot carbon dust
seen in other sources.

\begin{figure}
\setlength{\unitlength}{1mm}
\begin{center}
  \leavevmode
 \includegraphics[width=8cm]{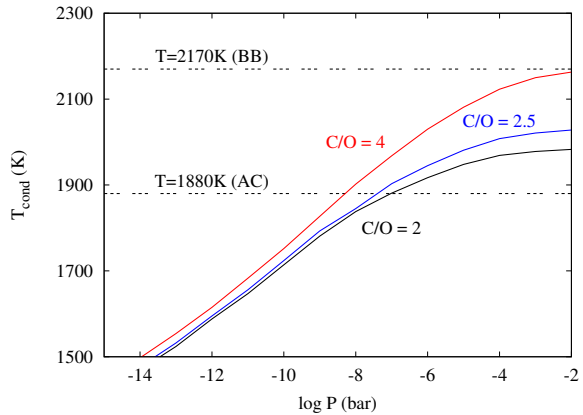}
   \caption[]{Condensation temperature for graphite grains for the 
   values of C/O indicated. Data from Table~2 of \cite{lodders95}.
   Data for blue curve (C/O $=2.5$) have been linearly interpolated 
   from Table~2 of \citeauthor{lodders95}. 
   Dotted horizontal lines 
   depict temperatures of hot blackbody and AC dust (Component~2),
   from Fig.~\ref{spec}b.\label{lodders}}
\end{center}
\end{figure}

While the stellar photosphere of SO was still visible, 
\cite{asplund99} determined the C/O ratio to be 
$\sim2.5$. The condensation sequence in these conditions has been 
considered by \cite{lodders95}, who found that the condensation 
temperature for {\em graphitic} 
\citep[as opposed to amorphous, as discussed here and in][]{evans20}
carbon in such an environment is very sensitive to the 
pressure. Their results for C/O $=2$ and 4, together with our own
linear interpolation for C/O $=2.5$, are shown in Fig.~\ref{lodders}. 
Allowing for possible differences between the condensation 
conditions for AC and graphite, and for uncertainties in the 
C/O ratio, it seems probable that AC could condense at 
$\sim1880$~K at pressure $\simeq10^{-7.5}$~bar),
requiring a density of carbon atoms
$\simeq{1.2\times10^6}$~cm$^{-3}$.

The 2020 data, combined with those from 2013, suggest
therefore that Component~2 represents the {\em very recent}
(since 2013 June) ejection of further AC-forming material.
If this is the case the dust mass is $\sim2.6\times10^{-14}$\Msun\ 
(again for the case of AC), far lower than the 
earliest dust mass deduced by \cite{evans20} ($\sim5\times10^{-10}$\Msun) 
following the 1998 dust ejection event.
Alternatively, Component~2 may be a small dust ``cloud'', as defined by
\cite{hinkle20}, produced by a single brief ejection event, and which
may not grow any further.

\cite{evans20} suggested that a dust component having temperature
437~K in 2014 March, and 411~K in 2016 July, pointed to renewed 
mass-loss and dust formation sometime in the period 2008--2014,
when the dust mass from this formation event was a few 
$\times10^{-8}$\Msun. The dust we see now at $624$~K (Fig.~\ref{spec}b) 
is very likely from the same ejection event that produced the 
550~K dust seen by \citeauthor{hinkle14} in 2013. This indicates
that there has been further mass-loss and dust formation
between 2013 and 2020. Our identification of Component~2 as 
yet another, more recent, dust formation event demonstrates that, 
as far as dust production is concerned, SO
continues to be active.
Further observations are required to verify whether Component~2
is indeed a post-2013 dust formation event, in which case the 
dust temperature will show clear evidence of declining.

We conclude that the stellar remnant remains unseen -- and that 
SO is still puffing out clouds of soot.

\subsection{The emission lines\label{emlines}}

\begin{figure}
\setlength{\unitlength}{1mm}
\begin{center}
  \leavevmode
 \includegraphics[width=8cm]{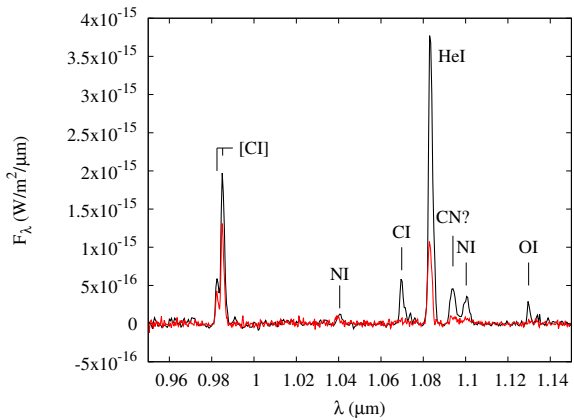}
  \caption[]{Difference between the emission spectrum 
   in 2013 \citep[red;][]{hinkle14} and in 2020 (black; this work) in 
   the 0.95--1.15\mic\ region. The continuum has been subtracted from
   both spectra to highlight the difference in the emission lines.
   \label{Sak_13-20-diff}}
\end{center}
\end{figure}

Numerous emission lines are present in the 2020 September
spectrum, some of which were reported by \cite{hinkle14}. With 
the exception of the \pion{He}{i} line at 2.059\mic, these lie 
in the 0.87--1.7\mic\ region; this portion of the spectrum is shown 
in Fig.~\ref{lines}. Each line has been fitted with a gaussian with 
respect to the adjacent linear continuum; the line 
centres, fluxes, and the proposed identifications
are given in Table~\ref{fluxes}. 
The line fluxes generally increased between 
2013 June and 2020 September, as illustrated for the 0.95--1.15\mic\
region in Fig.~\ref{Sak_13-20-diff}.

Most of the lines arise from transitions in
which the upper level is only a few eV above the ground; the only
exceptions are the \pion{He}{i} lines, which arise from levels
$\sim20$~eV above ground. This is significant in view of the 
expectation that the central star has an effective temperature 
$T_*\simeq10^5$~K and luminosity $L_*\sim10^{3.8}$\Lsun\ (see above).
{This is explored in Section~\ref{star} below.}

\subsubsection{The \pion{O}{i} lines\label{oi}} 

The \pion{O}{i} lines at 0.8448\mic\ ($^3$P$-^3$S$^o$) and
1.1290\mic\ ($^3$D$^o_3-^3$P$_2$) are both prominent in 
Fig.~\ref{lines}; however the 1.3168\mic\ line 
($^3$S$_1-^3$P$_2$) is not detected. In most stellar environments
this would suggest that continuum fluorescence is unlikely to play 
a significant role in the excitation of the first two of the above
\pion{O}{i} lines \citep[e.g.,][]{mathew18}. The 
7002\,\AA\ and 7254\,\AA\ \pion{O}{i} lines, which would also be expected if 
continuum fluorescence were important \citep{RA02}, were not present 
in an optical spectrum of SO obtained in 2013 \citep{vanhoof15a}. 
Also, if recombination were significant, the 7990\,\AA\ line, the lower
level of which is the upper level of the 0.8448\mic\ line, is
expected to be present; but this line was also absent in the 2013
optical spectrum \citep{vanhoof15a}. Although there is 
no near-simultaneous red-NIR spectroscopy, we are led to
conclude that continuum fluorescence and recombination are unlikely 
contributors to the excitation of \pion{O}{i}.

This leaves collisional excitation (by electrons or shocks) and 
Ly-$\beta$ (Bowen) fluorescence (which arises from a close coincidence 
between the wavelength of Ly-$\beta$, 1025.7222\,\AA\ and that of the 
$^3$P$-^3$D$^o$ transition in \pion{O}{i}, 1025.76\,\AA), as  
possibilities \citep[e.g.,][]{srivastava16}. 
However the presence of Ly-$\beta$ photons in a H-deficient 
environment is clearly problematic, and it is unlikely that such 
photons will have remained in the circumstellar environment 
since before the H-deficient phase.
While the role of Ly-$\beta$ in the case of SO might be taken 
by \pion{He}{ii} 6--2 ($\lambda=1025.273$\,\AA), the upper level of 
this transition is at 50~eV, in stark contrast to the relatively low 
excitation seen in the GNIRS spectrum (see Table~\ref{fluxes}). 
In any case there are no other \pion{He}{ii} lines in either the 
optical \citep{vanhoof15a} or the NIR spectrum presented here. 
Furthermore, there are no other plausible transitions in other ions that 
have wavelengths close to that of the $^3$P$-^3$D$^o$ transition in 
\pion{O}{i}.

We therefore consider collisional excitation to be the most 
likely excitation mechanism, but even this mechanism 
is not without difficulty. 
For electron excitation, the expected flux ratio 
$I(1.1290\mic)/I(0.8448\mic)$ has been calculated by 
\cite{bhatia95}, to be $\ltsimeq0.03$ for a wide range of 
electron temperatures (5000~K -- 100000~K) and densities 
($10^4$~cm$^{-3}$ -- $10^{12}$~cm$^{-3}$), far lower than the 
(dereddened) value in SO, $\simeq0.75$. On the other hand,
there may be some evidence for electron collisional excitation of
\fion{C}{i} lines (see Section~\ref{ci} below), and collisional
excitation of \pion{He}{i} 1.083\mic\ line in SO was suggested by
\cite{eyres99}. Collisional excitation
in shocks might also be a possibility, and has been invoked
by \cite{vanhoof07} to account for the decline in emission 
line fluxes (Section~\ref{ci}).

Previous investigations of
excitation mechanisms have been in the context of environments
in which abundances are, even if not solar, not too far removed 
therefrom. Excitation mechanisms need to be re-examined for environments
in which hydrogen is severely deficient.

\subsubsection{The \fion{C}{i} lines.\label{ci}} 

The \fion{C}{i} lines at 0.8729, 0.9824 and 0.9850\mic\ 
\citep[also reported by][]{vanhoof07,vanhoof15a} are
of particular interest. The two longest wavelength lines
\citep[also observed by]
[who did not report the 0.8729\mic\ line]{hinkle14}
originate from the same $^1$D$_2$ level and, combined with 
the shortest wavelength line, provide useful constraints on 
the emitting gas. 

At electron temperature $T_e=10^4$~K, the 
critical electron density above which the upper $^1$D$_2$ 
level is mostly collisionally, rather than radiatively, 
de-excited is $n_e=1.6\times10^4$~cm$^{-3}$ \citep{liu95}. 
\citeauthor{liu95} also find that the flux ratio 
$I(0.9824\mic+0.9850\mic)/I(0.8729\mic)$ is a potential pointer to the 
excitation mechanism; we note that this ratio is essentially 
independent of the assumed reddening as the lines are close
in wavelength. We find this flux ratio to have the value
$\simeq9.5$ in the 2020 September spectrum (Fig.~\ref{lines} and
Table~\ref{fluxes}). \cite{hinkle14} found fluxes of 
$5.9\times10^{-19}$~W~m$^{-2}$ and $2.4\times10^{-18}$~W~m$^{-2}$ 
in the 0.9824\mic\ and 0.9850\mic\ lines, respectively, in 2013 June; 
a re-examination of their 2013 spectrum suggests that the 
0.8729\mic\ line may be weakly present, with a flux of 
$2.1[\pm0.5]\times10^{-19}$~W~m$^{-2}$, but in any case
the upper limit on the flux ratio is $\sim14$.

Figure~2 of \cite{liu95} shows the parameter space for the 
flux ratio, $n_e$ and $T_e$. Assuming $T_e=10^4$~K, the above
two values of the flux ratio seem inconsistent with 
radiative recombination. On the other hand they seem consistent
with collisional excitation by electron impacts if the 
electron density increased from $\sim10^3$~cm$^{-3}$
in 2013 to $\sim10^4$~cm$^{-3}$ in 2020. 
The implied rise in $n_e$ may be linked to the increased
density in the region where the ``Component~2'' dust has
formed. Given the critical
density for collisional de-excitation from the $^1$D$_2$ level,
we might expect the 0.9824\mic\ and 0.9850\mic\ lines
to be quenched if $n_e$ continues to rise.
\cite{vanhoof07} have argued that the decline in the emission 
line fluxes is inconsistent with photo-ionisation,
but is consistent with excitation in a shock, of uncertain
origin, that occurred at some time prior to 2001, after which
the gas cooled and recombined.

\subsubsection{The \pion{He}{i} lines\label{hei}}

\begin{figure}
\setlength{\unitlength}{1mm}
\begin{center}
  \leavevmode
 \includegraphics[width=8cm]{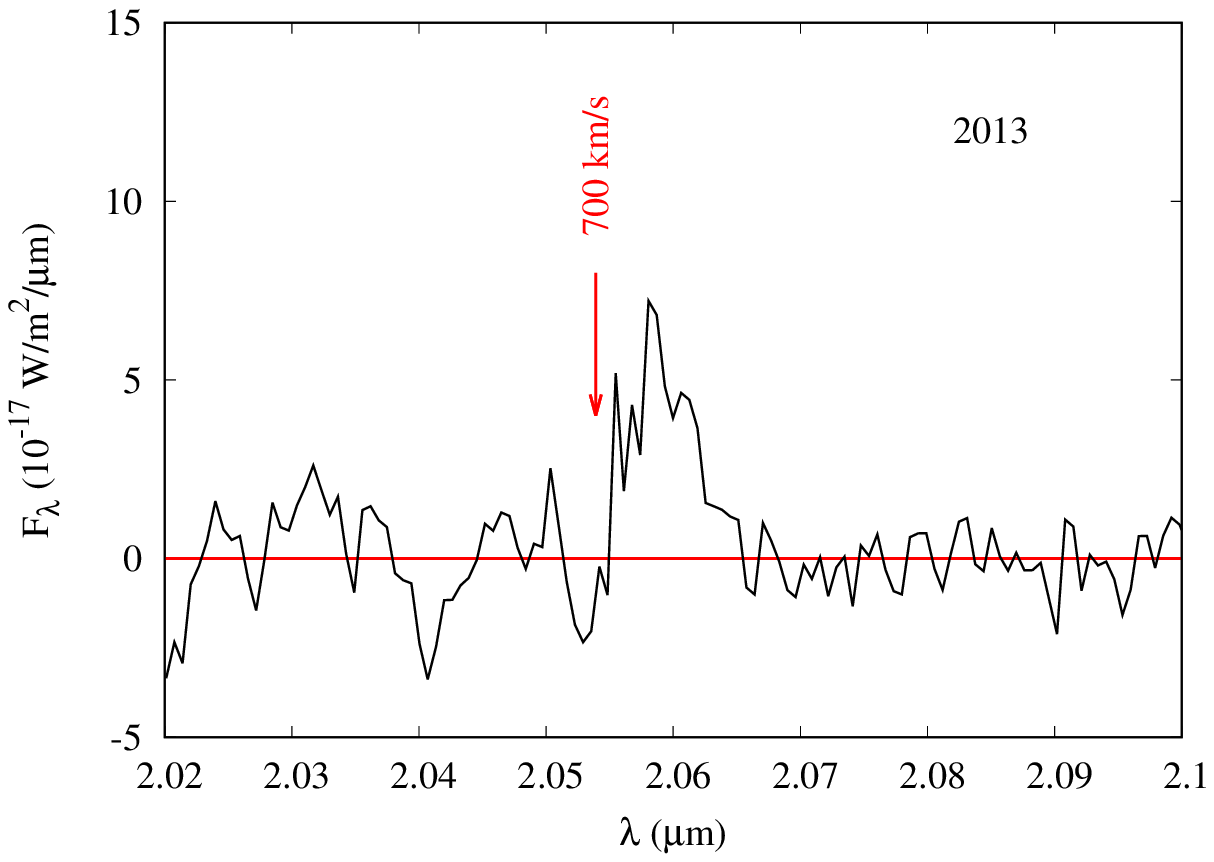}
  \includegraphics[width=8cm]{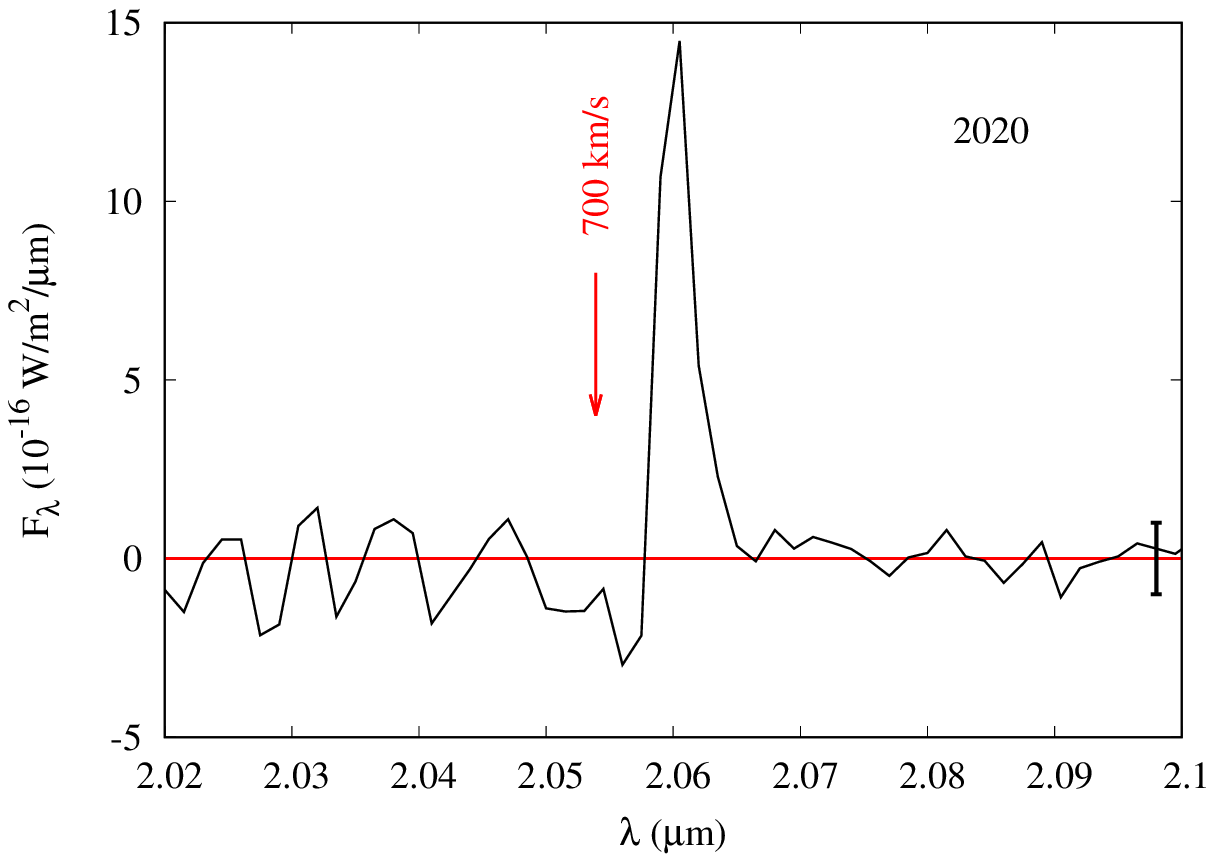}
\caption[]{Possible P~Cyg profiles in \pion{He}{i} 2.058\mic;
upper panel, 2013, lower panel 2020. The error bar in the 2020 
data, derived from the fluctuations in the data points in the
intervals 2.02--2.045\mic\ and 2.075--2.10\mic, is $\pm1\sigma$. 
Note that, at the resolution of this spectrum, the telluric 
transmission, which contains numerous strong and narrow absorption
lines of CO$_2$, does not resolve individual CO$_2$ lines from 
one another and that the observed CO$_2$ absorption reaches 
a maximum depth of only $\sim25$\% below the continuum.  \label{PCyg}}
\end{center}
\end{figure}

The \pion{He}{i} triplet at 1.0833\mic\ is
the strongest emission feature in the spectrum in 2013 and 2020,
and increased in strength by a factor of $\sim3.5$ between the 
two epochs (see Fig.~\ref{Sak_13-20-diff}). 
A weak P~Cygni profile in the \pion{He}{i} line at 
2.0589\mic\ appears to be present in both 2013 and 2020 spectra
(see Fig.~\ref{PCyg}); the absorption appears to be broad and 
centered at $\sim-700$\vunit. This value is similar
to those deduced for SO in mid-1998 (Evans et al., 
to be submitted), and of the same order as determined by
\cite{hinkle20} for the \fion{C}{i} lines (see Section~\ref{ci}).
There is no obvious P-Cygni profile in the 
\pion{He}{i} 1.083\mic\ line.
This is not unexpected as details of the 
radiative transfer in each of these two lines is different.

The dereddened flux ratio $I(1.0833\mic)/I(2.0589\mic)$ $\simeq32$.
This is close to the value \citep[30--33; see][]{benjamin99}
for a gas having electron temperature $10^4$~K and electron 
density in the range $10^4-10^6$~cm$^{-3}$; furthermore this 
line ratio excludes values of $n_e\ltsimeq10^4$~cm$^{-3}$.
This seems consistent with the $n_e$ values implied by
the \fion{C}{i} lines (see Section~\ref{ci}).
This suggests that the \pion{He}{i} lines
do not suffer significant circumstellar extinction, consistent
with a geometry in which the \pion{He}{i} emission arises in
jets that are perpendicular to the dust disc \citep{hinkle20}
and the lines of sight to the jets do not pass through the dust disc.

\subsubsection{Molecular features\label{mol}} 

\begin{figure}
\setlength{\unitlength}{1mm}
\begin{center}
  \leavevmode
 \includegraphics[width=8cm]{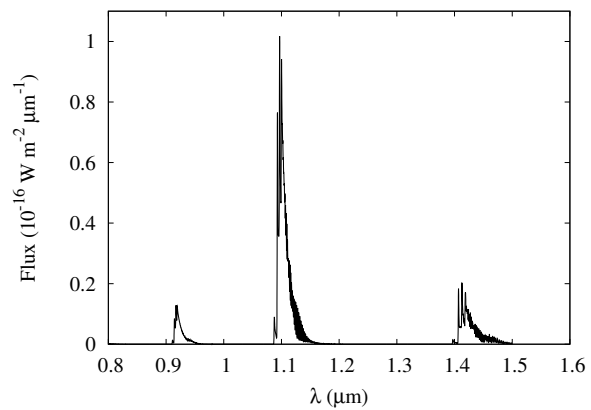}
   \caption[]{Calculated emission from CN at 1000~K, using CN data
   from \cite{brooke14}; main emission feature is at 1.09\mic.
   The theoretical CN spectrum has not been convolved
   with the instrumental resolution. See text for discussion.\label{CN_1000}}
\end{center}
\end{figure}

\cite{eyres98} reported NIR spectra of SO shortly after its VLTP. 
The \cite{BR} C$_2$ bands ($A'^2\Sigma^-_g-X'^3\Pi_u$)
at 1.4494\mic\ (1--2), 1.5063\mic\ (2--3), 1.7762\mic\ (0--0), 
1.8326\mic\ (1--1) were prominent in absorption, as were the first overtone 
CO bands. \cite{eyres98} also reported that the red system of the 
$A^2\Pi-X^2\Sigma$ electronic transition in CN was very prominent
in absorption at 1.1\mic\ ($\nu'=0\rightarrow\nu''=0$), and was 
the strongest of the CN bands in their spectral range.

Much later in the evolution of the 
VLTP, \cite{evans06} found the presence of small molecules (HCN, 
acetylene, polyynes) in the  mid-IR. Around the same time
(2003--2004) \cite{worters09} found that the CO fundamental was 
prominent in absorption against the dust shell. 

\cite{vanhoof18} 
reported a complex of some five emission lines around 0.9\mic, which 
had been emerging since 2013; they tentatively identified some of 
the latter features with the (1,0) and (0,0) transitions in 
the red system of the $A^2\Pi-X^2\Sigma$  electronic transition
in CN.

We find no evidence for emission or absorption by CO or 
C$_2$ over the entire wavelength range in the 2020 September observation.
The case for or against CN is less clear-cut. We note that, while all 
the stronger lines in the 0.8--2.5\mic\ range are robustly 
identified, there are prominent features that remain unidentified,
around 1.0940\mic\ and 1.4075\mic; these are tantalisingly close to
some of the CN $\Delta\upsilon = 0$ and $\Delta\upsilon = 1$ band-head
positions \citep{eyres98}. Indeed the 1.09\mic\ feature was tentatively 
identified with CN by \cite{hinkle20}.

Fig.~\ref{CN_1000} shows the expected 
emission from optically thin \nucl{12}{C}N in Local Thermal 
Equilibrium (LTE) at 1000~K, calculated using molecular data from 
\cite{brooke14}. However the situation is complicated by (a)~the likelihood
that the environment of SO is highly non-LTE
\citep*[see, e.g.,][for the effects of non-LTE on CO emission]{liu92},
(b)~the certain presence of \nucl{13}{C}N, in view of the low
\nucl{12}{C}/\nucl{13}{C} ratio, and the fact that optical depth effects
are not included in our simple model. A simple comparison is not 
feasible without detailed modelling along these lines. There are no other 
plausible candidates for these features.

An emission feature at 1.09\mic\ was detected in the NIR spectrum of
the peculiar helium nova V445~Pup by \cite{ashok03}, who attributed it
to \pion{C}{ii}; however other expected \pion{C}{ii} lines were not 
obviously present. This object, like SO, displayed numerous He
and \pion{C}{i} lines, but none of hydrogen 
\citep{ashok03}. Both SO and V445~Pup have C-rich, 
H-deficient environments.

\subsection{Where is the hot central star?\label{star}}

The evolutionary track given in \cite{hajduk05} 
suggests that SO's central star should, at the time of the 2020
observation, have had an effective temperature close to $10^5$~K. 
However, as noted in Section~\ref{emlines}, the emission lines -- 
with the exception of \pion{He}{i} -- arise in upper levels having
relatively low excitation. Moreover, all the identified
species are neutral.

This inevitably raises the question: is the expected hot remnant of 
the 1996 VLTP present?
One might reasonably expect a source at $10^5$~K to ionise the 
surrounding gas, but there is no evidence for a higher degree of
ionisation and excitation. Even at slightly lower temperatures,
for example $\sim30000-$50000~K (typical of the central star 
temperatures of a large number of planetary nebulae), a higher 
degree of excitation would be expected in the NIR spectrum 
\citep[e.g.,][]{rudy01}. A plausible explantion is that ionising 
radiation does not reach the emitting gas seen in Fig.~\ref{lines}
because of internal
extinction by the dust. However the dust geometry takes the 
form of a disc \citep{chesneau09}, and some hard radiation must emerge 
along the disc axis. Indeed, that there is emitting material, 
which is extended, along this axis is demonstrated by 
the observations of \cite{hinkle20} (see also Section~\ref{ext0} below): 
in the presence of a hot ($\sim10^5$~K) source, there would surely be 
highly ionised material, such as \pion{He}{ii}, in this region.

Another explanation is that material close to the 
central star, and located within the dust disc, is indeed highly 
ionised. However this ionised material would be visible 
at long wavelengths even through $\sim10$~mag of visual extinction.
For example, \cite{asplund99} determined the neon abundance in SO's
photosphere to be $\sim1.4$~dex higher than solar, so we would expect 
to see evidence of higher ionisation states of Ne, such as \fion{Ne}{ii}
(12.814\mic), \fion{Ne}{iii} (15.555\mic) and \fion{Ne}{v} (14.322\mic), 
at which wavelengths extinction by the dust would be much reduced. 
However none of these lines are present in {\it Spitzer} IRS 
and SOFIA spectra of SO \citep{evans06,evans20}, while the 
first two {\em are} present in the {\it Spitzer} IRS spectrum of 
the VLTP V605~Aql \citep{evans06}.

The implication is that there is no hot star;
at present the temperature of the ``photosphere'' of the central 
object of SO is much less than $10^5$~K, and that the evolutionary 
models, at least for SO, may need revisiting.

\section{Extended emission}
\label{ext0}

\begin{table}
\caption{Helicentric radial velocities of peak emission as determined from the 
\pion{He}{i} and \fion{C}{i} profiles in Fig.~\ref{extended}.
\label{compare2}}
 \begin{tabular}{cccccc} 
Offset            &  0\farcs90 & 0\farcs45 & 0\farcs0 & 0\farcs45 & 0\farcs90 \\
along slit $('')$ &   NNE       & NNE      & Centre   &  SSW      & SSW \\ \hline
\pion{He}{i} velocity (\vunit) & +321 & +210 & --50 & --125 & --111 \\
\fion{C}{i} velocity (\vunit) & +217 & +169 & --61 & --151 & --195 \\\hline
  \end{tabular}
\end{table}

\begin{figure*}
\setlength{\unitlength}{1mm}
\begin{center}
  \leavevmode
 \includegraphics[width=7cm]{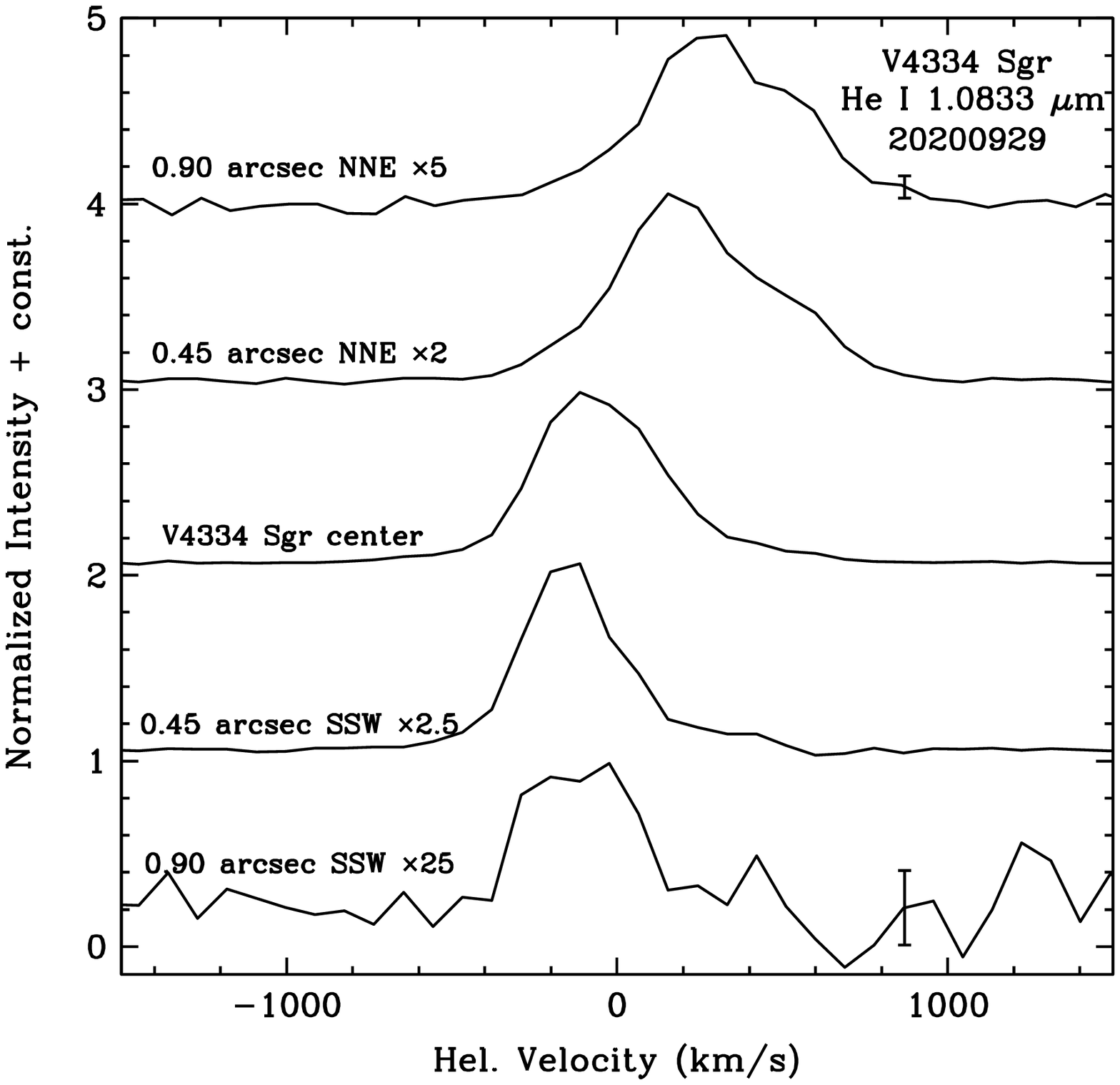}
  \includegraphics[width=7cm]{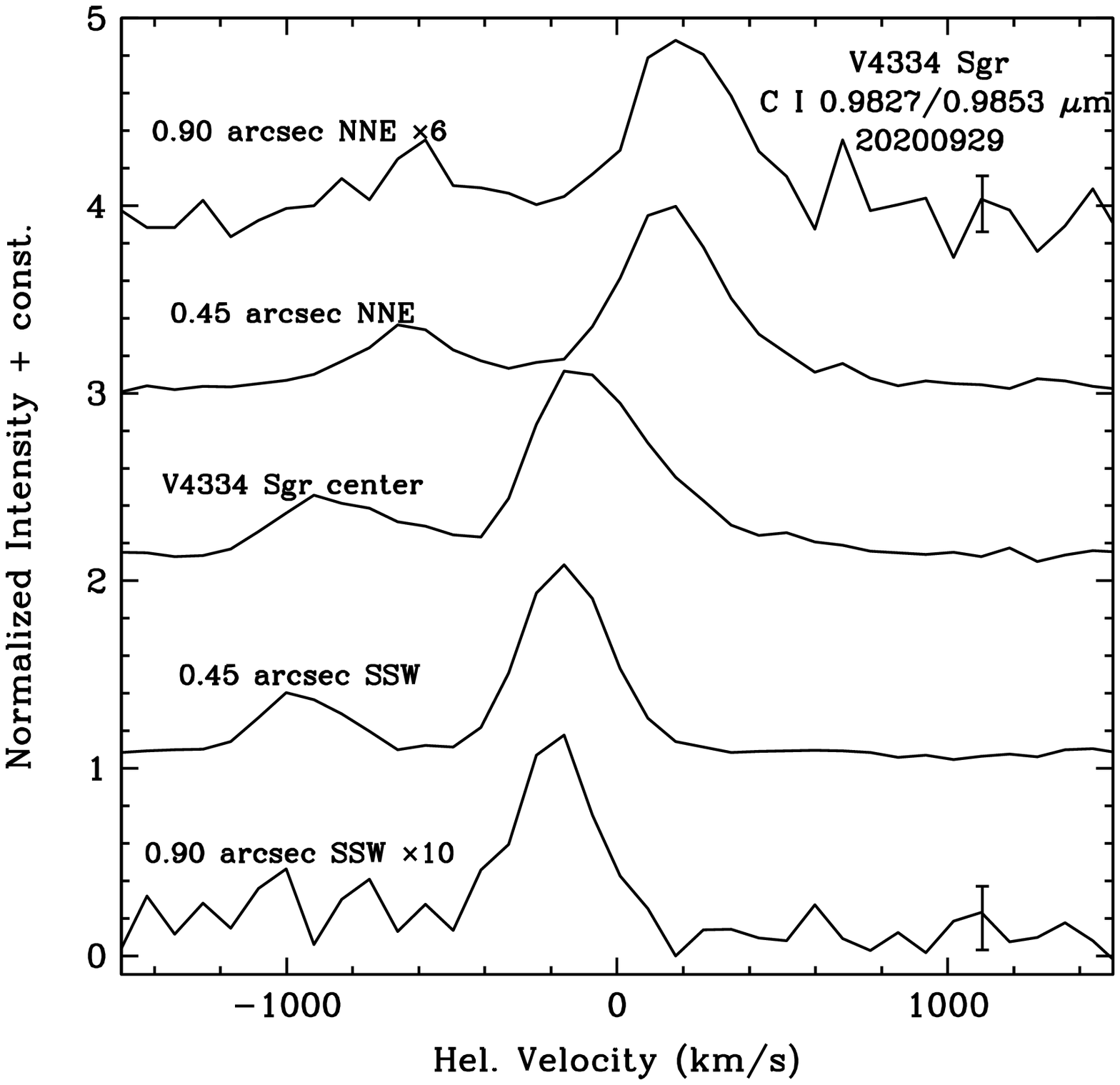}
  \includegraphics[width  =7cm]{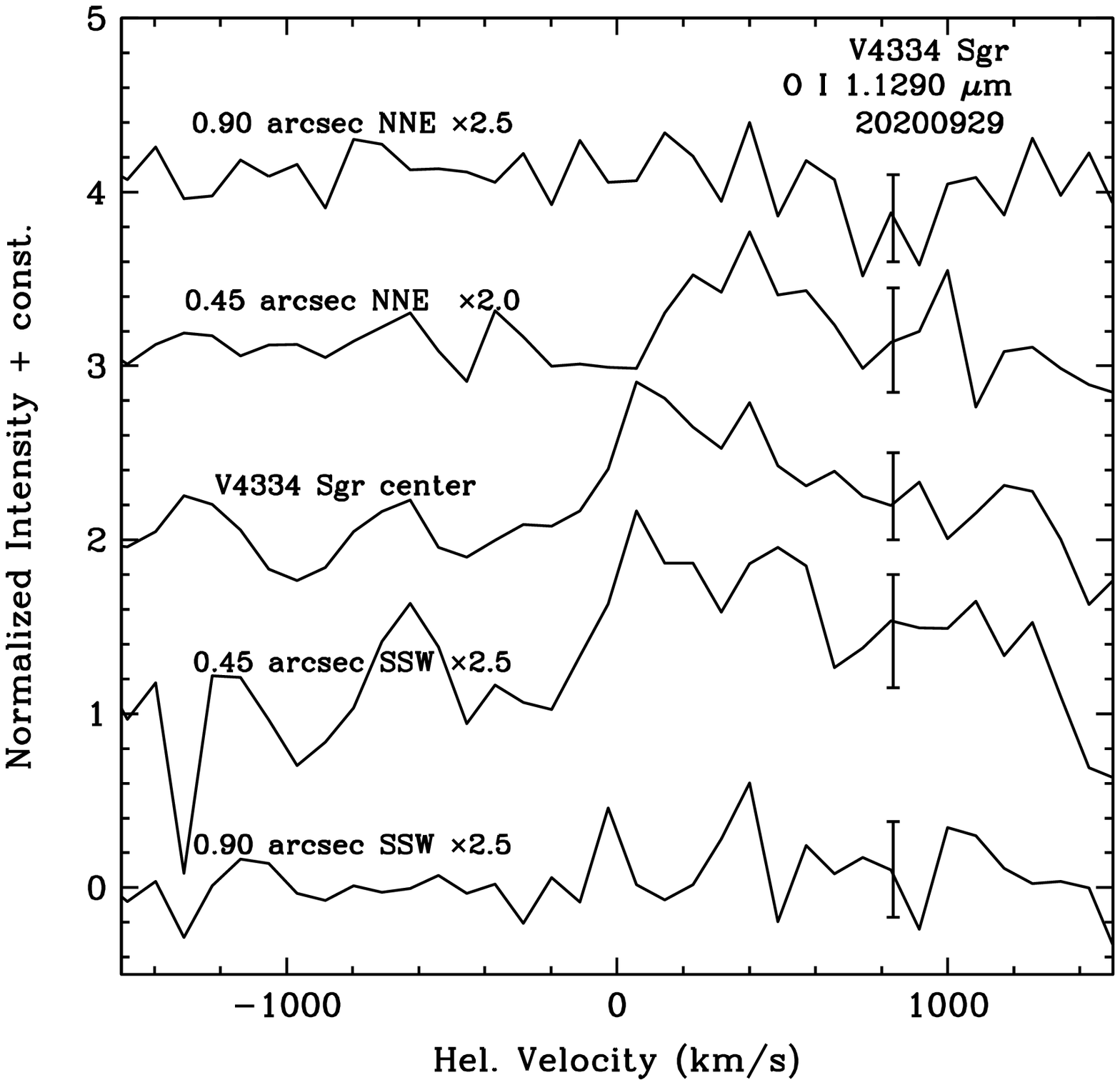}
    \includegraphics[width=7cm]{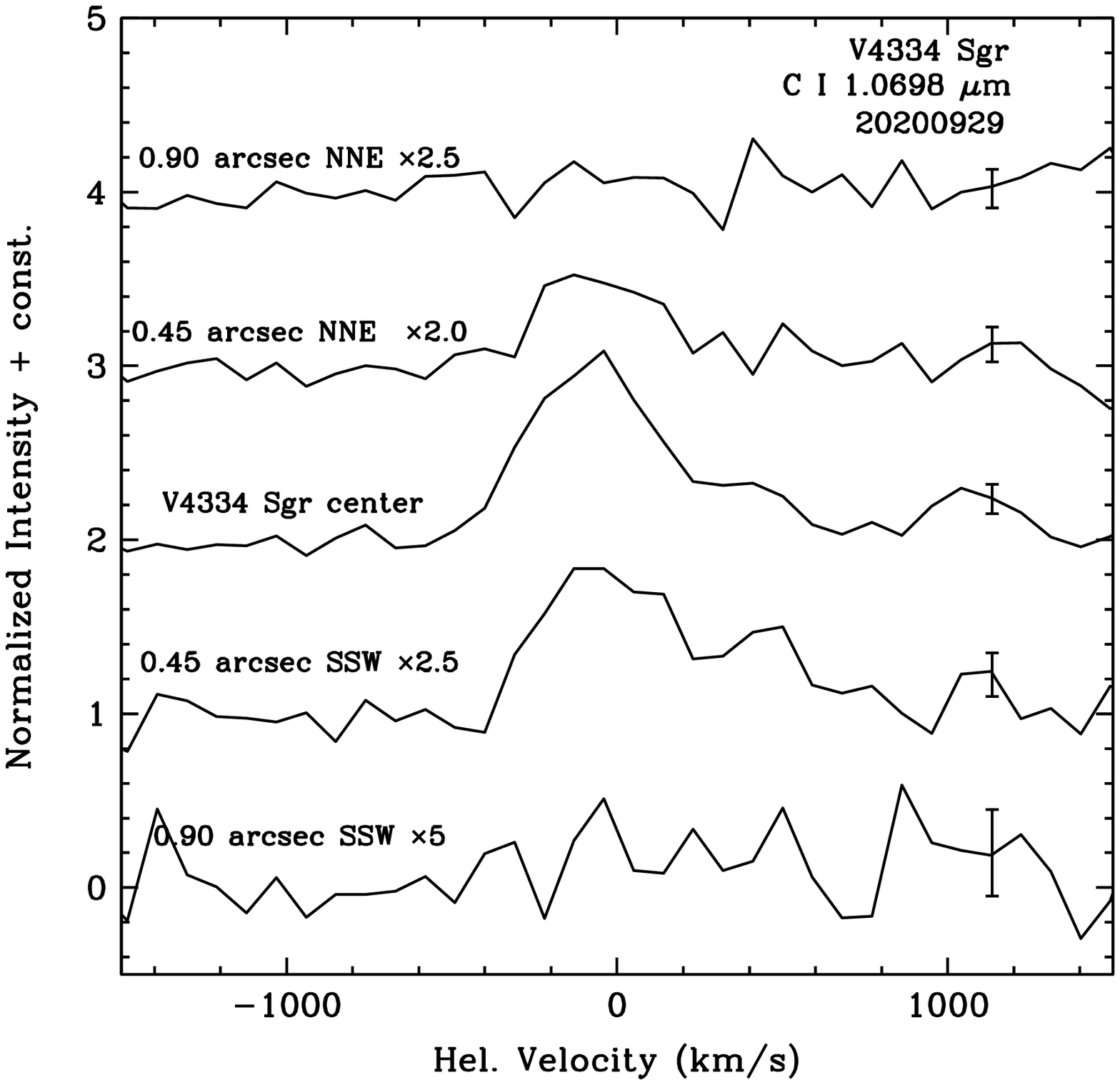}
   \caption[]{Top left: the \pion{He}{i} 1.0833\mic\ profile at five 
   positions along the  slit. 
   Top right: same for the \fion{C}{i} 0.9824/0.9850\mic\ lines;
   the heliocentric velocity in this case is that for the stronger
   0.9850\mic\ line; the shorter wavelength line appears as a 
   feature at $-796$\vunit. Each spectrum is the sum of three adjacent rows 
   (covering 0\farcs45). 
   Bottom left: same for \pion{O}{i} 1.1290\mic.
   Bottom right: same for \pion{C}{i} 1.0688\mic.
   The error bars are $\pm1\sigma$; the error bars are 
   negligible for some of the higher signal-to-noise spectra. 
   \label{extended}}
\end{center}
\end{figure*}

\cite{hinkle14} found that emission in the \pion{He}{i} 1.0833\mic\ line
in 2010 September had radial velocities in the
range $\sim-800$\vunit\ to $\sim+200$\vunit; they further found that 
the emission was extended, by about $\sim1\farcs9$ to zero intensity. 
They found that \fion{C}{i} 0.9850\mic\ was also extended, but that its
range of radial velocities was significantly different
($\sim-400$\vunit\ to $\sim+300$\vunit) from that of \pion{He}{i}. 
Recent observations \citep{hinkle20} show that the \pion{He}{i} 
emission region had expanded during the period 
2010--2019, and has a bipolar structure with a dark lane
at position angle $\simeq120^\circ$, consistent with that of the 
dense dusty disc detected by \cite{chesneau09}, and with that of the
old planetary nebula at whose centre SO is located.

Our 2020 September data (Fig.~\ref{extended}) also show that the 
\pion{He}{i} and \fion{C}{i} 0.9850\mic\ emission are extended. 
At 0\farcs9 NNE of the central star, the radial velocity of peak 
\pion{He}{i} emission peaks is $\sim+300$\vunit, but significant 
emission is present up to 
$\sim+900$\vunit. There is no evidence for emission beyond 
$\sim1''$ from the central object at this position angle. 
\cite{hinkle20} found that the \pion{He}{i} 1.0833\mic\ and
\fion{C}{i} 0.9850\mic\ lines were resolved in a 2015
spectrum, the linewidths in their 2015 spectrum implied
an expansion velocity of $\sim500$\vunit. 

The emission to the SSW is much 
weaker and is blueshifted, with the peak emission at heliocentric 
radial velocity of $\sim-200$\vunit. 
Similar emission profiles are seen in the \fion{C}{i} 0.9824/0.9850\mic\
lines (note that, in Fig.~\ref{extended}, the 0.9824\mic\ \fion{C}{i}
line appears as a feature $\sim800$\vunit\ blueward of the stronger 
0.9850\mic\ line).
We have determined the peak heliocentric velocities for each of the 
\pion{He}{i} 1.0833\mic\ and \fion{C}{i} 0.9850\mic\ lines in the 
top two panels of Fig.~\ref{extended}. 
The velocities of peak emission are given in Table~\ref{compare2}.

There is marginal evidence for extended emission, out to 0\farcs45, 
in the \pion{O}{i} line at 1.1290\mic, and in the \pion{C}{i}
1.0688\mic\ line. The signal-to-noise ratio
for these two features is rather low (see Fig.~\ref{extended}).

Although the signal-to-noise ratios in some of the line profiles are
rather low, Fig.~\ref{extended} suggests that the line profiles
of the \pion{He}{i}, \pion{O}{i} and \pion{C}{i} lines 
are broadly similar, indicating that they arise in the same 
region. This would be consistent with our conclusion 
(see Section~\ref{oi}) that they suffer similar extinctions.

\section{Conclusions}

We have presented a new 0.8--2.5\mic\ spectrum of the Very
Late Thermal Pulse object V4334~Sgr (Sakurai's Object). 
We conclude that:
\begin{enumerate}
\item {the effective temperature of the stellar remnant
is unlikely to be as high as that implied by current evolutionary models;}
\item the large increase in continuum flux density near 1\mic\
 is due to a very recent episode of extremely hot, likely amorphous 
 carbon, dust formation;
 \item the stellar component remains unseen;
 \item the \pion{He}{i} lines suffer negligible
 circumstellar extinction, suggesting that they arise in jets having
 lines of sight that do not pass through the dust disc;
 \item the relative intensities of the \fion{C}{i} NIR lines 
 suggest a rise in the electron density from 2013 to 2020;
 \item there is some evidence for emission by the CN radical; there
 is no evidence for emission by other molecules in the 0.82--2.5\mic\
 range.
\end{enumerate}

In view of the continuing rapid spectral evolution of Sakurai's 
Object, further IR observations of it in this and
other wavelength bands are strongly encouraged 

\section*{Acknowledgments}


This paper is based on observations obtained at the international 
Gemini Observatory, a program of NSF's NOIRLab, which is managed 
by the Association of Universities for Research in Astronomy (AURA) 
under a cooperative agreement with the National Science Foundation, 
on behalf of the Gemini Observatory partnership: 
the National Science Foundation (United States), 
National Research Council (Canada), 
Agencia Nacional de Investigaci\'{o}n y Desarrollo (Chile), 
Ministerio de Ciencia, Tecnolog\'{i}a e Innovaci\'{o}n (Argentina), 
Minist\'{e}rio da Ci\^{e}ncia, Tecnologia, 
Inova\c{c}\~{o}es e Comunica\c{c}\~{o}es (Brazil), 
and Korea Astronomy and Space Science Institute (Republic of Korea).

Data were also obtained under IRTF programs 2020A-001 and 
2020A-010. The Infrared Telescope Facility is operated by the
University of Hawaii under contract 80HGTR19D0030 with the 
National Aeronautics and Space Administration.

DPKB is supported by a CSIR Emeritus Scientist grant-in-aid and 
is being hosted by the Physical Research Laboratory, Ahmedabad. 
RDG was supported by the United States Airforce.
CEW was supported by USRA/SOFIA contract SOF 07-0027 (NASA prime award NNA17BF53C).

\section*{Data availability}
The raw data in this paper
that were obtained at the Gemini Observatory
are available from the Gemini Observatory
Archive, https://archive.gemini.edu/.

\bsp

\label{lastpage}

\end{document}